\documentclass[aps,prl,superscriptaddress,twocolumn,floatfix,showpacs]{revtex4}
\usepackage{graphicx}
\usepackage{amsmath}
\usepackage{bm}
\usepackage{color}

\begin{document}
	
	\title{Penetration of a supersonic particle at the interface in a binary complex plasma}
	
	\author{He Huang}
	\affiliation{College of Science, Donghua University, Shanghai 201620, PR China}
	\author{Mierk Schwabe}
	\author{Hubertus M. Thomas}
	\affiliation{Institut f\"ur Materialphysik im Weltraum, Deutsches Zentrum f\"ur Luft- und Raumfahrt (DLR), We{\ss}ling 82234, Germany}
    \author{Andrey M. Lipaev}
    \affiliation{Joint Institute for High Temperature, Moscow 125412, Russia}
    \affiliation{Moscow Institute of Physics and Technology (MIPT), Dolgoprudny 141701, Russia}
	\author{Cheng-Ran Du}
    \email{chengran.du@dhu.edu.cn}
    \affiliation{College of Science, Donghua University, Shanghai 201620, PR China}
    \affiliation{Member of Magnetic Confinement Fusion Research Centre, Ministry of Education, Shanghai 201620, PR China}
	
\begin{abstract}
	The penetration of a supersonic particle at the interface was studied in a binary complex plasma. Inspired by the experiments performed in the PK-3 Plus Laboratory on board the International Space Station, Langevin dynamics simulations were carried out. The evolution of Mach cone at the interface  was observed, where a kink of the lateral wake front was observed at the interface. By comparing the evolution of axial and radial velocity, we show that the interface solitary wave is non-linear. The dependence of the background particle dynamics in the vicinity of the interface on the penetration direction reveals that the disparity of the mobility may be the cause of various interface effects.
\end{abstract}

	
\maketitle

A complex plasma is a weakly ionized gas containing small solid particles \cite{Fortov_2005,Morfill_2009}. The particles are highly charged by collecting ions and electrons. Using video microscopy, localized structures and dynamics can be directly recorded in the experiments. Various phenomena such as formation of crystal lattice \cite{Chu_1994,Thomas_1994}, wave phenomena \cite{LinI_2018,Schwabe_2007,Merlino_2014,Bandyopadhyay_2008,Thomas_2006,Williams_2014}, and instabilities \cite{Schwabe_2009,Couedel_2010} can be studied in complex plasmas. A binary complex plasma contains two types of microparticles of different sizes, which can either be mixed \cite{Huang_2019,Du_2019} or form a phase separated system \cite{Wysocki_2010,Jiang_2011,Du_2012}. It was discovered that phase separation can still occur due to the different force balance for microparticles of different sizes under microgravity conditions despite the criteria of spinodal decomposition not being fulfilled \cite{Sutterlin_2009,Killer_2016}. An interface between separated phases emerges and various interfacial phenomena are investigated \cite{Yang_2017,Sun_2018}.

Recently, wakes excited by a moving disturbance in complex plasmas have attracted much attention \cite{Dubin_2000,Zhdanov_2016,Nosenko_2007,Du_2014,Khrapak_2019,Zhukhovitskii_2015} since the first theoretical predictions \cite{Havnes_1995,Havnes_1996}. The disturbance can be imposed by either a laser beam \cite{Melzer_2000,Nosenko_2002} or extra particles \cite{Jiang_2009,Caliebe_2011,Schwabe_2017}. If the disturbance moves faster than the sound speed in the complex plasma, the wakes exhibit a V-shaped structure in two-dimensional (2D) case \cite{Samsonov_1999,Samsonov_2000} and a conical structure in three-dimensional (3D) case \cite{Jiang_2009,Schwabe_2011,Caliebe_2011}, known as Mach cone. In the ground laboratory, the particles are levitated in the (pre)sheath and form a 2D plasma crystal in the case of strong coupling conditions. The extra particles can travel either above or below the particle layer, exciting a repulsive or attractive Mach cone, respectively \cite{Samsonov_1999,Du_2012b}. Under microgravity conditions, the particles form a relatively homogeneous 3D complex plasma. The penetration of the extra particles results in a moving disturbance inside the particle cloud, generating a 3D Mach cone if moving faster than the speed of sound \cite{Jiang_2009,Schwabe_2011,Zhukhovitskii_2012,Zaehringer_2018}.

\begin{figure}[!ht]
	\includegraphics[width=20pc]{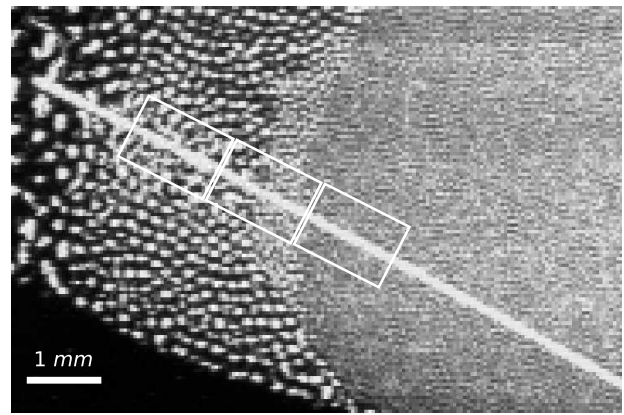}
	\caption{Snapshot of the penetration of an extra particle across the interface of a binary complex plasma in an experiment. Thirteen consecutive images were overlaid, where the trajectory of the penetrating particle was shown as a straight white line. Three locations are highlighted by the rectangles, corresponding to the insets in Fig.~\ref{fig2}(a-c).}
	\label{fig1}
\end{figure}

\begin{figure*}[!htbp]
    \includegraphics[width=38pc]{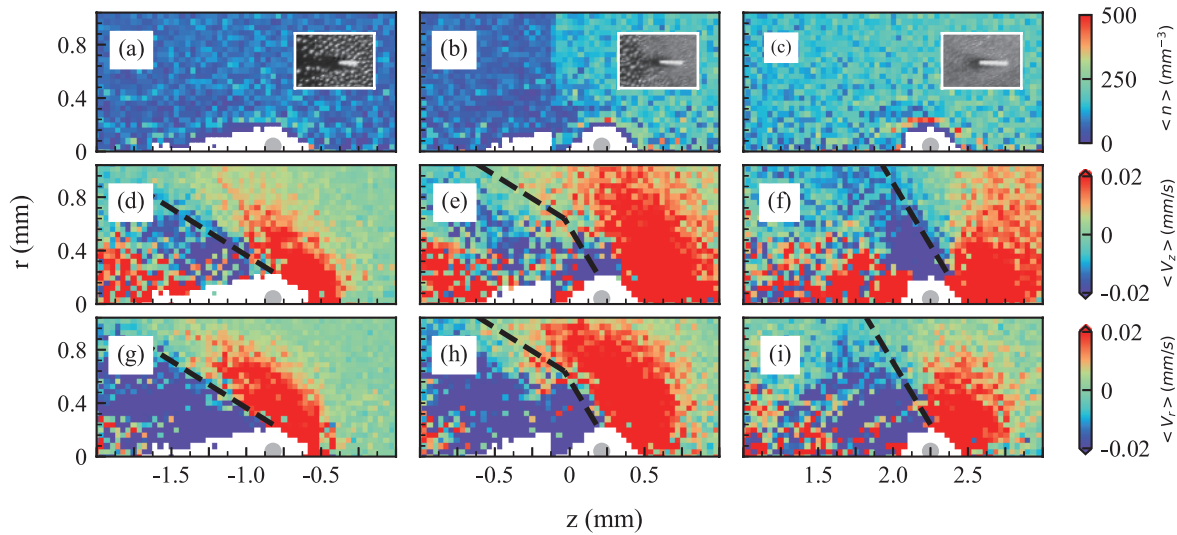}\hspace{1pc}%
    \begin{minipage}[h]{38pc}\caption{\label{fig2} Distribution of the density (a-c) and the velocity in z-direction (d-f) and in radial direction (g-i) of the background particles in a binary complex plasma in the Langevin dynamics simulation. The penetration velocity of the extra particle (shown as a grey semicircle) is set as $50$~mm/s. The left, middle and right panels correspond to the moment where the extra particle is in the big particle cloud, in the vicinity of the interface, and in the small particle cloud, respectively. For comparison, the experiment images at a similar moment are shown in the inset of (a-c), correspondingly. The cone structure is highlighted by dashed lines in (d-i).  }
    \end{minipage}
\end{figure*}

In this paper, we present a numerical simulation to study the evolution of a lateral wake excited by a supersonic extra particle at the interface, inspired by an experiment observation under microgravity conditions. The experiment was performed in the PK-3 Plus Laboratory on board the International Space Station (ISS). Technical details of the setup can be found in the Ref.~\cite{Thomas_2008}. A neon plasma was produced by a capacitively-coupled radio-frequency (rf) generator in push-pull mode at $13.56$~MHz. The binary complex plasma was prepared by injecting two types of particles. The first type is melamine formaldehyde (MF) particles of a diameter of $3.42$~$\mu$m, while the second type is SiO$_{2}$ particles of a diameter of $1.55$~$\mu$m. In addition, agglomerated larger particles were present on the outside of the particle cloud. Using video microscopy \cite{Thomas_2008}, a cross section of particle cloud (illuminated by a laser sheet) was recorded at a rate of $50$~frame-per-second (fps). The gas pressure was set at $20$~Pa, and the discharge voltage was set at $20$~V.

As we can see in Fig.~\ref{fig1}, the particles of two types were phase-separated, mainly due to the difference of the ion drag force \cite{Killer_2016}. The small particles were confined on the right side, while the big particles were on the left side. The extra particle moved from the left to the right, leaving a straight trajectory across the interface. The driving force is still not fully understood \cite{Schwabe_2011}. It might be due to a rocket force acting on the extra particle \cite{Nosenko_2010}.

We employed the python library $aircv$ based on the SIFT feature detector algorithm \cite{Lowe_2004} to track the penetrating particle. It accelerated from $20$~mm/s to $60$~mm/s, and the velocity near the interface was ${{\sim}50}$~mm/s \footnote{The acceleration can also be revealed by the prolongation of the shape of the extra particle in the snapshot.}. As we see in the insets of Fig.~\ref{fig2}(a-c), a cavity around the supersonic particle appeared. The Mach cone structure of the lateral wake emerged as the extra particle crossed the interface. The structure became evident as its velocity reached ${{\sim}60}$~mm/s, where the cone angle decreased dramatically.

However, due to the high density of the background particles, it is difficult to obtain the full details of individual particles in the cross section. In order to study the dynamics of this phenomenon quantitatively, we performed Langevin dynamics simulations, where the positions and velocities of each particle can be acquired easily \cite{Hou_2009,Schwabe_2013}. The equation of motion including damping from the neutral gas is given as
\begin{equation}
\label{eq1}
  m_{i}\ddot{r}_{i} +m_{i}\nu_{i}\dot{r}_{i}=-\sum_{j\ne i}\bigtriangledown\phi_{ij} +F_{id,i}+F_{c,i}+L_{i},
\end{equation}
where $r_{i}$ is the position of the particle $i$, $m_{i}$ is the mass, $\nu_{i}$ is the damping rate, and $L_{i}$ is the Langvin force. The Langevin force is defined by $\langle L_{i}(t)\rangle=0$ and $\langle L_i(t) L_i(t+\tau)\rangle=2\nu_{i}m_{i} k_B T\delta(\tau) I$, where $T$ is the temperature of the heat bath, $\delta(\tau)$ is the delta function, and $I$ is the unit matrix. The particles interact with each other via the Yukawa potential,
\begin{equation}
\label{eq2}
  \phi_{ij}= \frac{Q_{i} Q_{j}}{4 \pi \epsilon_{0}r_{ij}}exp(-\frac{r_{ij}}{\lambda}),
\end{equation}
where $\lambda$ is the Debye length, $Q_{i}$ is the charge of particle $i$ and $Q_{j}$ the charge of a neighboring particle $j$, separated by interparticle distance $r_{ij}$. The particle cloud was confined by the ion drag force $F_{id,i}$ directed in the negative z direction, and the confinement force $F_{c,i}(=-\nabla{\Phi}Q_{i})$ resulted from the confinement potential $\Phi$. The ion drag force was assumed to be constant for small particles and for big particles, respectively. The confinement potential was assumed to be parabolic, i.e., ${\Phi}=1/2 C z^{2}$, with a constant $C$. As result, the particle cloud was phase separated with the small particles located to the right of the big particles, as we see in Fig.~\ref{fig1}. The Langevin dynamics simulations were performed with $4000$ small particles and $1500$ big particles, using LAMMPS in NVE ensemble \cite{Plimpton_1995}. The rest of the parameters were set as in Table~\ref{tab1} \cite{Thomas_2008,Schwabe_2018,Yang_2017}. Note that the simulation was performed in the Cartesian coordinate system, while the analysis was performed in cylindrical coordinates, considering the symmetry of the system.

\begin{table}
\caption{\label{tab1} Parameters in the Langevin dynamics simulation: microparticle mass $m$, charge $Q$, damping coefficient $\nu$, ion drag force $F_i$, strength of confinement potential $C$ and Debye length $\lambda$}.
\begin{ruledtabular}
\begin{tabular}{ccccccc}
   type     & $m$  & $Q$ & $\nu$       & $F_{i}$ & $C$       & $\lambda$ \\
            & [kg] & [e]   & [s$^{-1}$]  & [fN]  & [V/m$^2$] & [$\mu$m]  \\
   \hline
   MF      & $3.3\times10^{-14}$ & $\phantom{0}6000$    & $50.7$ & $63$ & $10^4$ & $100$ \\
   SiO$_2$ & $3.6\times10^{-15}$ & $\phantom{0}2700$    & $91.3$ & $18$ & $10^4$ & $100$ \\
   extra   & -                   & $30000$              & -      & -    & -      & $100$
\end{tabular}
\end{ruledtabular}
\end{table}

Despite the fact that the extra particle accelerated along its trajectory in the experiment, we set the velocity of the extra particle as constant in the simulation for simplicity. To focus on the effect of the extra particle on the interface, we select the penetration velocity as $v_{c}=50$~mm/s. The density distribution is shown in Fig.~\ref{fig2}(a-c). Particle-free cavities are clearly seen around the extra particle, which are caused by its repulsion on the surrounding particles \cite{Khrapak_2019}. The cavity in the big particle cloud has a more elongated shape than that in the small particle cloud. However, it is truncated into two separated parts at the interface. This remarkable feature is caused by the fact that the small particles are much more mobile than the big particles so that they fill in the cavity at the interface before the big particles in the rear do. The radial and the axial velocity distribution of the background particles are shown in Fig.~\ref{fig2}(b-d) and (g-i), respectively. Clearly, the sound speed in the big particle cloud is smaller than that in the small particle cloud, which results in a smaller angle of the Mach cone. The lateral wake front is kinked at the interface, highlighted by the dashed lines in Fig.~\ref{fig2}(e,h).

\begin{figure}[!ht]
	\includegraphics[width=20pc]{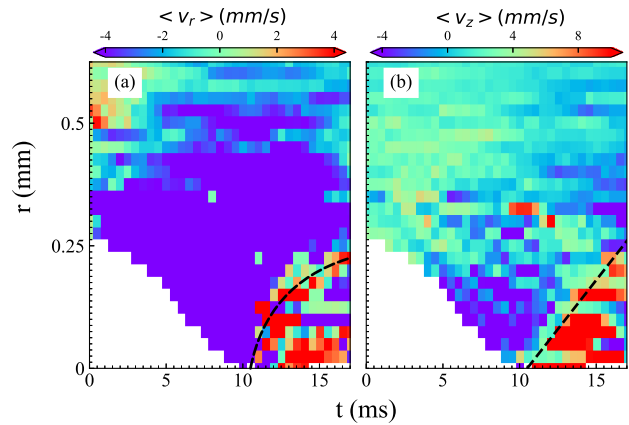}
	\centering
	\caption{ The evolution of the solitary wave by particle radial velocity (a) and axial velocity (b) at the interface. The cavity is left blank. The wave front is highlighted by a black dashed line.}
	\label{fig3}
\end{figure}

The penetration of the extra particle at the interface excited a solitary wave, propagating along the interface. We select all small particles within $50$~$\mu$m from the interface in the simulation and obtain the evolution of the axial and radial velocity. For better resolution, the simulation was repeated $20$ times with different initial conditions, and the results were averaged. The results are shown in the periodgram in Fig.~\ref{fig3}. As the extra particle left the interface, the cavity shrank instantaneously. The small particles moved towards the territory of the big particles, shown by the blue fraction in Fig.~\ref{fig2}(b). As soon as the cavity was closed, a highly dissipative solitary wave was excited, propagating along the interface outwards. The propagation of this solitary wave was revealed by both the evolution of the radial and axial velocity, however, at a different pace. While the evolution of the axial velocity shows that the wave propagated at a constant speed, the evolution of the radial velocity shows the propagation slowed down. This implies the non-linearity of this solitary wave, whose propagation range is below $500$~$\mu$m before it was dissipated.

\begin{figure}[!ht]
	\includegraphics[width=18pc]{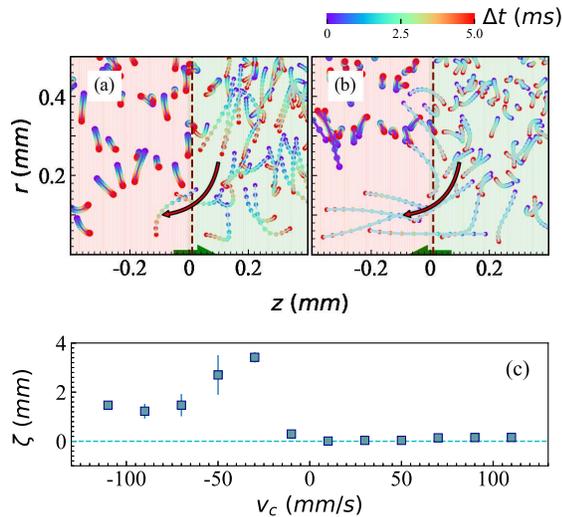}%
	\centering
	\caption{(a,b) Trajectories of background particles for two cases where the penetration directions of the extra particle are opposite. The green arrows represent the penetration direction of the extra particles. The colored curves represent the trajectories of particles at different times $\Delta t$, counting from the moment that the extra particle crossed the interface with a distance of $0.5$~mm. Thicker lines represent big particles while the thinner lines represent small particles. The red-dashed line is the position of the interface.  (c) The dependence of the penetration depth $\zeta$ of small particles on the speed of the extra particle $v_c$, where the interface position is marked by a dashed line.}
	\label{fig4}
\end{figure}

It is interesting to study the dependence of the reaction of the background particles in the vicinity of the interface on the penetration direction and speed. For this purpose, we performed a series of additional simulations in which we varied the speed and direction of the extra particle. We define the penetration depth $\zeta$ as the maximal distance of the small particles entering the large particle cloud from the interface. A positive speed $v_{c} > 0$ means that the extra particle moves from the cloud of big particles into that of small particles. As we can see in Fig.~\ref{fig4}(c), for positive penetration speeds, the small particles barely crossed the interface and intruded into the territory of the big particles. In contrast, for negative speeds, the small particles followed the extra particle, filled the cavity, and intruded into the territory of the big particles. At $v_{c}\approx-30$~mm/s, the small particles reached the deepest depth. However, when the speed of the penetrating particles became even faster, the depth decreased and finally reached saturation. This is caused by the differences of the particles on the two sides of the interface, in terms of the mass, damping rate, as well as the strength of the interaction. These factors lead to the disparity of the particle mobility so that the small particles react with greater magnitude than big particles to the disturbances in the stimulations. In fact, this may be the essential cause of various interface effects such as reflection of waves at interfaces.

In conclusion, we performed Langevin dynamics simulations to study the penetration of a supersonic particle at an interface in a binary complex plasma. The Mach cone structure was observed, where a kink emerged at the interface. By studying the dependence of the reaction of the background particles in the vicinity of the interface on the penetration direction and speed, we show that the disparity of the mobility may be the cause of various interface effects.

The authors acknowledge support from the National Natural Science Foundation of China (NSFC), Grant No. 11975073. The PK-3 Plus project was funded by the space agency of the Deutsches Zentrum f\"ur Luft- und Raumfahrt e.V. with funds from the Federal Ministry for Economy and Technology according to a resolution of the Deutscher Bundestag under Grant No. 50WP1203. The authors acknowledge Roscosmos provided the PK-3 Plus laboratory launch and operation onboard of ISS. We would like to thank V. Nosenko for carefully checking the manuscript.

\end{document}